\documentclass[12pt]{article}
\usepackage{graphicx}
\renewcommand{\theequation}{\thesection.\arabic{equation}}

\newlength{\extraspace}
\newlength{\extraspaces}
\newcommand{\be}{\begin{equation}
\addtolength{\abovedisplayskip}{\extraspaces}
\addtolength{\belowdisplayskip}{\extraspaces}
\addtolength{\abovedisplayshortskip}{\extraspace}
\addtolength{\belowdisplayshortskip}{\extraspace}}
\newcommand{\ee}{\end{equation}}

\newcommand{\ba}{\begin{eqnarray}
\addtolength{\abovedisplayskip}{\extraspaces}
\addtolength{\belowdisplayskip}{\extraspaces}
\addtolength{\abovedisplayshortskip}{\extraspace}
\addtolength{\belowdisplayshortskip}{\extraspace}}
\newcommand{\ea}{\end{eqnarray}}

\newcommand{\newsection}[1]{
\vspace{12mm}
\pagebreak[3]
\addtocounter{section}{1}
\setcounter{equation}{0}
\setcounter{subsection}{0}
\setcounter{footnote}{0}
\noindent{\bf \thesection. #1}
\nopagebreak
\medskip
\nopagebreak}

\newcommand{\newsubsection}[1]{
\vspace{0.8cm}
\pagebreak[3]
\addtocounter{subsection}{1}
\noindent{\it \thesubsection. #1}
\nopagebreak
\vspace{2mm}
\nopagebreak}

\newcommand{\newappendix}[1]{
\vspace{12mm}
\pagebreak[3]
\addtocounter{section}{1}
\setcounter{equation}{0}
\setcounter{subsection}{0}
\setcounter{footnote}{0}
\noindent{\bf Appendix \thesection #1}
\nopagebreak
\medskip
\nopagebreak}

\newcounter{saveeqn}
\newcommand{\alpheqn}{\setcounter{saveeqn}{\value{equation}}%
 \stepcounter{saveeqn}\setcounter{equation}{0}%
 \renewcommand{\theequation}
     {\mbox{\thesection.\arabic{saveeqn}\alph{equation}}}}
\newcommand{\reseteqn}{\setcounter{equation}{\value{saveeqn}}%
  \renewcommand{\theequation}{\thesection.\arabic{equation}}}

\begin{document}
\addtolength{\baselineskip}{1.5mm}

\thispagestyle{empty}
\begin{flushright}
gr-qc/0206066\\
\end{flushright}
\vbox{}
\vspace{2.5cm}

\begin{center}
{\LARGE{Quantum interest in two dimensions
        }}\\[16mm]
{Edward Teo~~and~~K.~F. Wong}
\\[6mm]
{\it Department of Physics,
National University of Singapore, 
Singapore 119260}\\[15mm]

\end{center}
\vspace{2cm}

\centerline{\bf Abstract}\bigskip
\noindent
The quantum interest conjecture of Ford and Roman asserts that any
negative-energy pulse must necessarily be followed by an over-compensating
positive-energy one within a certain maximum time delay. Furthermore, the
minimum amount of over-compensation increases with the separation between 
the pulses. In this paper, we first study the case of a negative-energy 
square pulse followed by a positive-energy one for a minimally coupled,
massless scalar field in two-dimensional Minkowski space. We obtain 
explicit expressions for the maximum time delay and the amount of 
over-compensation needed, using a previously developed eigenvalue approach. 
These results are then used to give a proof of the quantum interest 
conjecture for massless scalar fields in two dimensions, valid for 
general energy distributions. 


\newpage

\newsection{Introduction}

Most forms of classical matter obey the weak energy condition (WEC) \cite{HE}.
This means that the local energy density $\rho(t)$, as measured by an
observer with proper time $t$, is always non-negative. The WEC has a 
number of important implications in general relativity; for example, 
it can be shown that matter obeying the WEC will necessarily form a 
singularity after a certain critical stage in gravitational collapse
\cite{Penrose,HE}.  On the other hand, matter violating the WEC would 
lead to a host of exotic predictions, such as the existence of traversable 
wormholes \cite{MT} and time machines \cite{MTY}, 
naked singularities \cite{Ford:1990,Ford:1992}, and `faster-than-light' 
travel \cite{Alcubierre,Olum}.

As it turns out, the WEC is violated in quantum field theory. A 
classic result of Epstein, et al.~\cite{Epstein} states that the 
(renormalised) energy density of a quantum field at a given space-time 
point can be arbitrarily negative. However, there are constraints coming 
in when the energy density is averaged over time. One such constraint 
is the averaged weak energy condition (AWEC), which states that the 
integral of the energy density over the observer's geodesic is 
non-negative \cite{Tipler}:
\be
\int_{-\infty}^\infty \rho(t)\,{\rm d}t\geq0\,.
\ee
The AWEC is known to hold for a number of quantum fields in Minkowski space
\cite{Klinkhammer}.

More recently, a new class of constraints, known as {\it quantum 
inequalities\/} (QI's), was derived by Ford and Roman 
\cite{Ford,Ford:1994,Ford:1996}. QI's provide lower bounds on the 
weighted average of the energy density seen by a geodesic observer:
\be
\int_{-\infty}^\infty\rho(t)f(t)\,{\rm d}t\geq \rho_{\rm min}\,,
\ee
where the weighting is supplied by a `sampling function' $f(t)$, i.e., 
a peaked function of time with unit integral and a certain characteristic 
width $t_0$. The lower bound $\rho_{\rm min}$ is a negative quantity that
depends on a number of parameters, including the type of quantum field,
the space-time it is defined on, and the sampling function used. 
For example, the QI for minimally coupled, massless scalar fields 
in $2n$-dimensional Minkowski space takes the general form \cite{Flanagan,FE}
\be
\label{QI_equation}
\int_{-\infty}^\infty\rho(t)f(t)\,{\rm d}t\geq
-\frac{1}{c_n}\int_{-\infty}^\infty\bigg|\bigg(\frac{\rm d}{{\rm d}t}\bigg)^n
f^{1/2}(t)\bigg|^2\,{\rm d}t\,,
\ee
where $c_n$ is a positive constant given by
\be
c_n\equiv\left\{\begin{array}{ll} 6 \pi  &\quad n=1; \\
n\pi^{n-1/2} 2^{2n} \Gamma (n-\frac{1}{2}) &\quad n\geq2.
\end{array} \right.
\ee
QI's have also been derived for a number of other cases 
\cite{Pfenning:1996,Pfenning:1997,Pfenning,FT1,Fewster:1999,Vollick:2000,Fewster:2001,Pfenning:2001}.

Now, for example, if the Lorentzian sampling function 
$f(t)=t_0/[\pi(t^2+t_0^2)]$ is used in (\ref{QI_equation}), 
it can be shown that $\rho_{\rm min}\propto-t_0^{-2n}$. Thus, in the 
infinite sampling time limit, $\rho_{\rm min}\rightarrow0^-$, and so the QI 
reduces to the AWEC. This implies, in particular, that any negative-energy 
distribution must necessarily be followed (or preceded) in time by 
at least the same amount of positive energy. Ford and Roman 
\cite{Ford:1992,Ford:1999} have likened this to bank loans: the 
negative energies are loans, and the positive parts repayments. The 
AWEC then becomes the statement that all loans must be repaid in full
over time.

Ford and Roman have investigated this phenomenon for the case of
$\delta$-function pulses in two- and four-dimensional massless scalar 
field theory \cite{Ford:1999}, for a specific choice of sampling function. 
They showed that a negative $\delta$-function pulse must necessarily 
be followed by a positive one within a certain maximum time delay.
Furthermore, they found that the positive $\delta$-function pulse must 
{\it over\/}-compensate the negative one, by an amount which monotonically
increases with the separation between the pulses. Ford and Roman have 
conjectured that this is, in fact, a general phenomenon of quantum fields, 
and have termed it the {\it quantum interest conjecture\/}: the negative 
energy loan must be repaid with a positive interest `rate'; moreover 
there is a maximum term for the loan. 

It is of obvious interest to check the quantum interest conjecture
for more general energy distributions and sampling functions. 
Pretorius \cite{Pretorius} has managed to extend the results in
the former direction. However, as with Ford and Roman's analysis,
a specific one-parameter family of sampling functions was assumed, and 
the appropriate QI optimised over it. This is, unfortunately, not 
guaranteed to give the best possible bounds for the interest `rate' and 
maximal loan term.

Fewster and one of the present authors \cite{FT2} have recently developed
a new approach to quantum interest for massless scalar fields in even 
dimensions, which involves turning it into an eigenvalue problem familiar 
from quantum mechanics. This is actually quite easy to see if we gloss over 
some technical issues (which are rigorously addressed in \cite{FT2}). 
If we write $\psi(t)=f^{1/2}(t)$ and integrate by parts $n$ times, 
(\ref{QI_equation}) can be recast in the form
\be
\label{positivity}
\langle\psi|H^{(n)}\psi\rangle\geq0\,,
\ee
where $\langle\,\cdot\,|\,\cdot\,\rangle$ is the usual $L^2$-inner 
product, and $H^{(n)}$ is the differential operator
\be
\label{H}
H^{(n)}\equiv(-1)^n\bigg(\frac{\rm d}{{\rm d}t}\bigg)^{2n} + c_n\rho(t)\,.
\ee
The condition (\ref{positivity}) for all $|\psi\rangle$ is equivalent to 
the requirement that $H^{(n)}$ does {\it not\/} have any negative eigenvalues. 

Hence, in this approach, the energy density $\rho(t)$ becomes (up to a 
constant) the potential of a generalised Schr\"odinger operator $H^{(n)}$, 
and the non-existence of negative eigenvalues to $H^{(n)}$ is equivalent 
to the QI being satisfied for $\rho(t)$. Note that the foregoing condition 
has to hold for all $\psi(t)$. Any $H^{(n)}$ obeying this condition is 
therefore guaranteed to give a $\rho(t)$ satisfying (\ref{QI_equation}) 
for {\it all\/} sampling functions $f(t)$, and we shall refer to it as a 
{\it QI-compatible\/} energy density, following \cite{FT2}. 

In two-dimensional Minkowski space, it happens that the operator 
$H^{(1)}=-{{\rm d}^2}/{{\rm d}t^2} + 6\pi\rho(t)$ has precisely the form
of the Hamiltonian in quantum mechanics. This will allow us to turn
the problem of quantum interest in two dimensions into a one-dimensional 
quantum-mechanical problem, where time is now the spatial coordinate. 
The negative parts of $\rho(t)$ become potential wells, and the positive
parts potential barriers. Given a form of the energy distribution, one 
can in principle find out whether bound states (with negative eigenvalues) 
exist for this potential, and therefore whether the energy distribution 
is QI-compatible or not.

The example of double $\delta$-function pulses was treated in \cite{FT2}
using this approach, albeit rather briefly, and bounds on the maximal 
loan term and interest `rate' were obtained. These bounds are tighter 
than those obtained in \cite{Ford:1999}, since they are automatically 
optimised over all possible sampling functions, and are limited only by 
how optimal the original QI used is. As it fortuitously turns out, the 
QI (\ref{QI_equation}) {\it is\/} optimal in two dimensions \cite{Flanagan}.

In the first part of this paper, we shall consider another example, 
namely that of double square pulses, and show in detail how the 
eigenvalue approach can be used to extract optimal bounds on the 
maximal loan term and interest `rate'. This case turns out to exhibit 
a much richer behaviour than the $\delta$-function pulse case, yet it 
is simple enough to be solved analytically. 

Of course, it would be desirable to eventually move beyond specific
examples, and see how the eigenvalue approach can be used to prove 
the quantum interest conjecture in general. A first step in this direction 
was made in \cite{FT2}, where a very general but non-constructive proof 
of the existence of maximal loan terms in any even dimension 
was given. However, the question of whether there is always
a positive interest `rate' was left open.

In the second part of this paper, a proof of the quantum interest 
conjecture for {\it general\/} energy distributions in two-dimensional
massless scalar field theory is given. Firstly, we explain how a theorem of 
Simon \cite{Simon}, regarding the existence of negative eigenvalues to 
the Hamiltonian, can be used to prove the existence of quantum interest 
for general energy distributions. We will then use our previous results 
on square pulses, together with the min-max principle \cite{RS-IV}, to 
prove that the interest owed increases with the term of the loan. 
Finally, we use the same method to give an alternative proof of the 
existence of a maximum loan term, and furthermore, obtain
an upper bound for this loan term.

\newsection{Double square pulses}

In this section, we shall investigate the case of square
pulses in detail using the eigenvalue approach of \cite{FT2}. 
In Sec.~2.1, we obtain a bound on the maximal separation between the 
negative square pulse and the compensating positive pulse (which need not
be square). Then in Sec.~2.2, we determine the minimum quantum
interest `rate' required in the case when the positive pulse is also square.

\newsubsection{Pulse separation}

Consider a negative-energy square pulse starting at time $t=0$, of 
magnitude $\rho_1$ and duration $a$. Let us suppose it is followed by 
a compensating positive pulse starting at time $T\geq a$, which may be of 
any shape. We wish to find the necessary conditions for this system to be
QI-compatible, which would, in particular, give us an upper bound on the
time separation $T-a$ between the two pulses. This involves constructing
the corresponding quantum-mechanical potential $V(t)\equiv6\pi\rho(t)$ 
and finding the necessary conditions for the non-existence of bound states
to this potential.

Now, consider the new potential obtained if the positive pulse is replaced
by an infinite wall of positive energy at time $T$. By the min-max principle
\cite{RS-IV}, the non-existence of bound states to the original potential
implies the non-existence of bound states to the new potential. We 
therefore turn our attention to this new potential. It is sketched
in Fig.~1, where we have set $V_1\equiv6\pi \rho_1$. The potential is 
divided along the time axis into four regions, I to IV, as indicated. We 
now find the conditions for the existence of normalisable 
bound-state `wavefunctions' $\Psi$ to the Schr\"odinger equation
\be
\label{Schrodinger}
H^{(1)}\Psi=-k^2\Psi\,,
\ee
with negative energy eigenvalue $-k^2$. 

\begin{figure}[t]
\begin{center}
\includegraphics{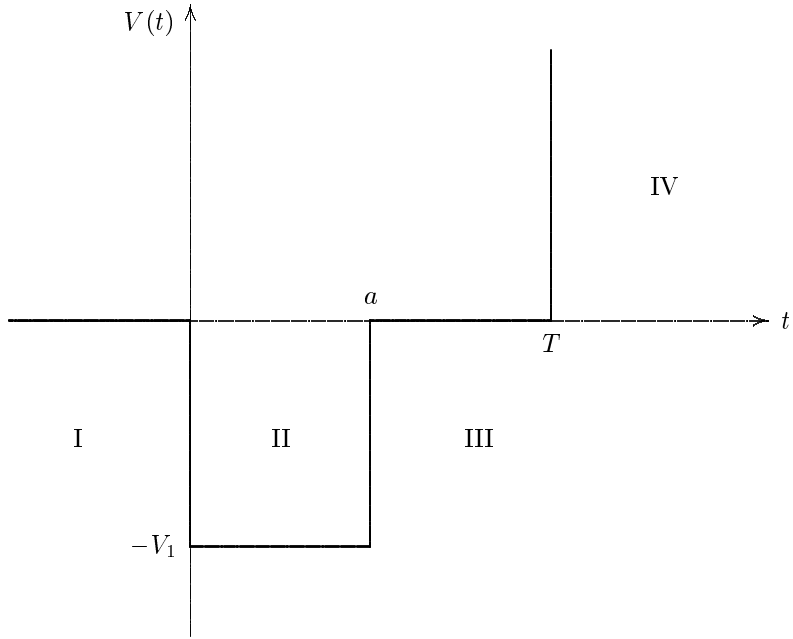}
\caption{Quantum-mechanical potential corresponding to a negative-energy
square pulse followed by an infinite positive-energy wall.}
\end{center}
\end{figure}

In region I, a solution which vanishes at infinity is 
\be
\label{Psi1}
\Psi_{\rm I}(t)={\rm e}^{kt},
\ee
where $k$ is assumed to be positive. Note that this solution 
is defined up to a normalisation constant which has not been included, 
as it will not affect the following arguments. Matching (\ref{Psi1}) 
across the boundaries into regions II and III in the standard way, we 
obtain the respective solutions:
\be
\Psi_{\rm II}(t)=\cos\omega t + \frac{k}{\omega}\sin\omega t\,,
\ee
and
\be
\Psi_{\rm III}(t) = \Big( \cos \omega a + \frac{k}{\omega} \sin \omega a \Big) 
\cosh k(t-a) + \Big( \cos \omega a - \frac{\omega}{k} \sin \omega a \Big) 
\sinh k(t-a)\,,
\ee
where $\omega\equiv\sqrt{V_1-k^2}$. For bound states to occur, 
we require $k<\sqrt{V_1}$ (see, e.g., \cite{GP}).

The infinite potential at $t = T$ implies that $\Psi_{\rm III}(t)$ has to
vanish at this point. This condition is equivalent to
\be
T-a=\frac{1}{k} \tanh^{-1} F(k)\,,
\label{SOLUTION_equation}
\ee
where 
\be
F(k)\equiv\frac{k}{\omega} \ \frac{ \cos \omega a + \frac{k}{\omega} \sin \omega a }
{ \sin \omega a - \frac{k}{\omega} \cos \omega a }\,.
\ee
Since the left-hand side of (\ref{SOLUTION_equation}) is non-negative, we 
must have $0\leq F(k)<1$.

\begin{figure}[t]
\begin{center}
\includegraphics{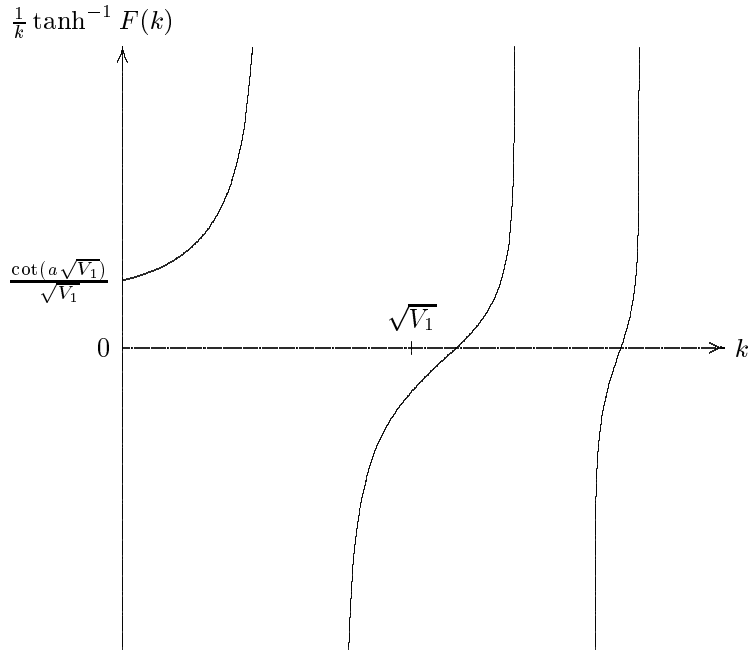}
\caption{Example of a graph of $\frac{1}{k}\tanh^{-1} F(k)$ against $k$.
The requirement for the non-existence of bound states is that the branches
do not intersect the $k$-axis in the interval $(0,\sqrt{V_1})$, as shown.}
\end{center}
\end{figure}

The requirement for QI-compatibility is precisely that there is {\it no\/} 
$k$ which satisfies (\ref{SOLUTION_equation}); we now work out 
the conditions for this to be so. An example of a graph of 
$\frac{1}{k}\tanh^{-1} F(k)$ is shown in Fig.~2. It consists of disjoint 
branches (similar to that of $\tan k$, for example), each of which is 
monotonically increasing. The proof of the latter fact, being somewhat 
technical, is relegated to Appendix A. Thus, a necessary condition for the 
non-existence of a solution $k\in(0,\sqrt{V_1})$ to 
(\ref{SOLUTION_equation}) is
\ba
T-a&\leq&\lim_{k=0}\left(\frac{1}{k}\tanh^{-1} F(k)\right)\cr
&=&\frac{\cot \left(a\sqrt{V_1} \right)}{\sqrt{V_1}}\,.
\label{PULSESEPB_equation}
\ea

Furthermore, we need to ensure that a solution does not come
from the {\it second\/} branch of $\frac{1}{k}\tanh^{-1} F(k)$.
This is equivalent to demanding that at no point in the region
$(0,\sqrt{V_1})$ does $\frac{1}{k}\tanh^{-1} F(k)$ vanish, which
translates to the requirement that
\be
\label{tanrhoneq}
\tan\omega a \neq -\frac{\omega}{k}\,,
\ee
for $0<\omega<\sqrt{V_1}$. It is straightforward to solve (\ref{tanrhoneq}) 
graphically: plotting the graphs of $\tan\omega a$ and 
$-\frac{\omega}{\sqrt{V_1-\omega^2}}$ together against $\omega$, it is 
clear that they do not intersect if and only if
\be
\label{AREAB_equation}
a \sqrt{V_1} \leq \frac{\pi}{2}\,.
\ee
As a consistency check, note that (\ref{AREAB_equation}) ensures that 
the right-hand side of (\ref{PULSESEPB_equation}) is always non-negative.

Thus, we have found that demanding QI-compatibility of the 
negative-energy square pulse imposes the two necessary conditions 
(\ref{AREAB_equation}) and (\ref{PULSESEPB_equation}),
which can respectively be written as
\be
a \sqrt{\rho_1} \leq \sqrt{\frac{\pi}{24}} \,,
\label{AREA_equation}
\ee
\be
T-a \leq \frac{ \cot \left( a \sqrt{6 \pi \rho_1} \right) }
{\sqrt{ 6 \pi \rho_1} }\,.
\label{PULSESEP_equation}
\ee
The first condition can be interpreted as a constraint on the
magnitude and duration of the negative square pulse: the more
negative it is, the shorter its duration must be. Such a result
is to be expected from the quantum inequalities. We recall that there 
is, however, no corresponding result for the case of a negative
$\delta$-function pulse \cite{FT2}. Again, this is not surprising
because the quantum inequalities do not place any restrictions on 
how negative a pulse can be, if it is localised to a definite point 
in time.

The second condition (\ref{PULSESEP_equation}) places an upper bound
on the time separation $T-a$ between the negative square pulse and the
subsequent positive one. Note that $T-a$ depends inversely on $\rho_1$ and
also on $a$: the more negative the square pulse is, or the longer
its duration is, the shorter the time interval must be before the
compensating positive pulse arrives. This is entirely consistent 
with the notion of quantum interest.

Finally, note that if we take the limit $a\rightarrow0$ such that
$B\equiv a\rho_1$ remains finite, the condition (\ref{PULSESEP_equation}) 
gives the maximum time separation
\be
T_{\rm max} = \frac{1}{6\pi B}\,.
\ee
This agrees with the corresponding result obtained in \cite{FT2}
for a negative $\delta$-function pulse of magnitude $B$.

\newsubsection{Quantum interest `rate'}

We now turn to the question of the quantum interest `rate', i.e., by how 
much must a positive square pulse {\it over\/}-compensate the negative one. 
To study this, we have to replace the infinite wall of Sec.~2.1 by a 
positive square pulse of magnitude $\rho_2$, starting at time $T$ and 
lasting for a duration $b$. We then find the necessary and sufficient 
conditions for this system to be QI-compatible. The corresponding potential
$V(t)$ is shown in Fig.~3, where as usual, $V_2\equiv6\pi \rho_2$.
There are now five different regions in time to consider when solving 
the Schr\"odinger equation (\ref{Schrodinger}).

\begin{figure}[t]
\begin{center}
\includegraphics{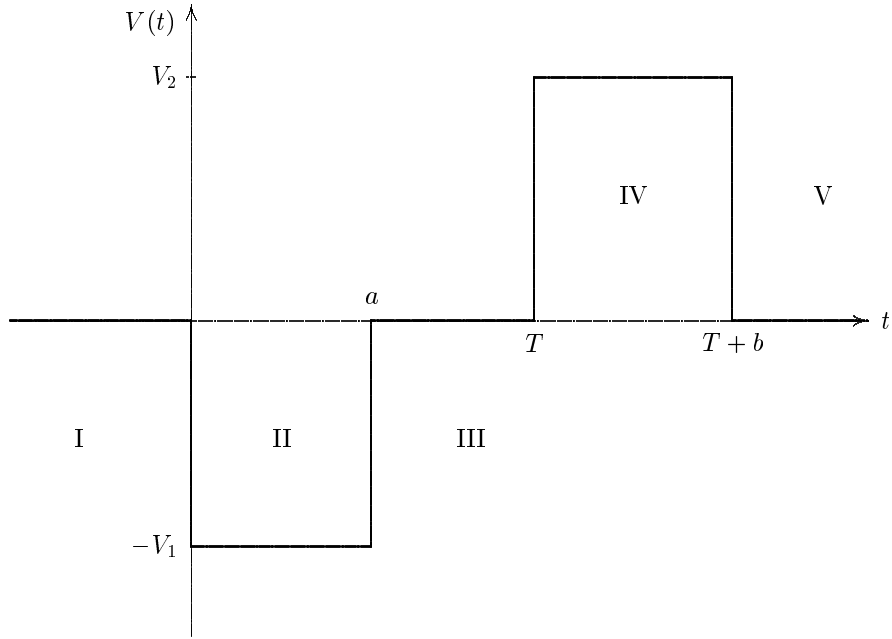}
\caption{Quantum-mechanical potential corresponding to the case of 
double square pulses.}
\end{center}
\end{figure}

The solutions in regions I, II and III are the same as those derived
in Sec.~2.1. Matching the solution $\Psi_{\rm III}(t)$ across the 
boundary $t = T$ into region IV gives
\be
\Psi_{\rm IV}(t)= C \cosh \lambda (t-T) + C' \sinh \lambda (t-T)\,,
\ee
where we have defined the constants $\lambda\equiv\sqrt{V_2+k^2}$, and 
\alpheqn
\ba
C &\equiv& \Big( \cos \omega a + \frac{k}{\omega} \sin \omega a \Big)
\cosh k(T-a)
+\left( \cos \omega a - \frac{\omega}{k} \sin \omega a \right)
\sinh k(T-a)\,, \\
C' &\equiv& \frac{k}{\lambda}\left[\Big( \cos \omega a + \frac{k}{\omega} \sin \omega a \Big)
\sinh k(T-a)
+\left( \cos \omega a - \frac{\omega}{k} \sin \omega a \right)
\cosh k(T-a)\right].~~~~~~~ 
\ea
\reseteqn
Similarly, matching this solution across the boundary $t = T+b$ into 
region V gives
\be
\Psi_{\rm V}(t)=D \cosh k (t-T-b) + D' \sinh k (t-T-b)\,,
\ee
where
\be
D \equiv C \cosh\lambda b+C'\sinh\lambda b\,, \qquad
D' \equiv \frac{\lambda}{k}\,(C \sinh \lambda b +C'\cosh\lambda b)\,.
\ee

Now, to prevent $\Psi_{\rm V}(t)$ from blowing up as $t\rightarrow\infty$,
we require $D+D'=0$. We obtain, after some calculation,
\ba
\label{DD}
D+D'&=&\left(\omega\sin\omega a-k\cos\omega a\right)g_{1}(k)G(k)\,,
\ea
where we have defined
\be
\label{Gk}
G(k)\equiv\frac{1}{k} \left( F(k) - \frac{g_{2}(k)}{g_{1}(k)} \right),
\ee
and
\alpheqn
\ba
g_{1}(k) &\equiv& \left(\frac{\lambda}{k} \cosh k(T-a) + \frac{k}{\lambda} \sinh k(T-a)\right) \sinh \lambda b + {\rm e}^{k(T-a)}\cosh \lambda b\,, \\
g_{2}(k) &\equiv& \left(\frac{\lambda}{k} \sinh k(T-a) + \frac{k}{\lambda} \cosh k(T-a)\right) \sinh \lambda b + {\rm e}^{k(T-a)}\cosh \lambda b\,.
\ea
\reseteqn
Note that both $g_{1}(k)$ and $g_{2}(k)$ are manifestly positive functions.

For QI-compatibility, we require that normalisable bound-state 
wavefunctions to (\ref{Schrodinger}) do {\it not\/} exist. In other words, 
we need to find the conditions under which $D+D'\neq0$ for all 
$k\in(0,\sqrt{V_1})$. Now, it can be seen that the right-hand side of 
(\ref{DD}) vanishes if and only if $G(k)$ vanishes. In Appendix B, it is 
shown that $G(k)>G(0)$ for all $k\in(0,\sqrt{V_1})$. Thus, we have 
QI-compatibility if and only if $G(0)$ is non-negative. Since,
\be
\label{lim_G}
\lim_{k=0}G(k)=\frac{\cot(a\sqrt{V_1})}{\sqrt{V_1}} 
- \frac{\coth(b\sqrt{V_2})}{\sqrt{V_2}} - (T-a)\,,
\ee
this means that
\ba
T-a &\leq& \frac{\cot(a\sqrt{V_1})}{\sqrt{V_1}} 
- \frac{\coth(b\sqrt{V_2})}{\sqrt{V_2}}\cr
&=& \frac{\cot(a\sqrt{6\pi \rho_1})}{\sqrt{6\pi \rho_1}} 
- \frac{\coth(b\sqrt{6\pi \rho_2})}{\sqrt{6\pi \rho_2}}
\,.
\label{INTEREST1_equation}
\ea

The constraint (\ref{INTEREST1_equation}) is clearly a generalisation 
of (\ref{PULSESEP_equation}). However, it now plays two roles: not only
does it provide an upper bound for the time separation between the two
pulses, it can also tell us how much the positive pulse must
over-compensate the negative one if it is rewritten in the form 
\be
\label{INTEREST2_equation}
\sqrt{V_2}\tanh(b\sqrt{V_2})\geq\left(\frac{\cot(a\sqrt{V_1})}{\sqrt{V_1}}
-(T-a)\right)^{-1}.
\ee
Let us highlight some implications of this constraint:

\begin{enumerate}

\item Since $x < \tan x$ and $\tanh x < x$ for $0<x<\frac{\pi}{2}$, we may
deduce, from (\ref{INTEREST1_equation}), the following sequence of
inequalities:
\be
a V_1 < \sqrt{V_1} \tan (a\sqrt{V_1}) \leq \sqrt{V_2} \tanh (b\sqrt{V_2}) 
< b V_2.
\label{TANTANH_equation}
\ee
This implies that
\be
a\rho_1 < b \rho_2\,,
\ee
i.e., the area (in the sense of magnitude multiplied by duration) of the 
positive pulse is strictly greater than the area of the negative pulse. 

\item In the limit when the two areas coincide, it follows from 
(\ref{TANTANH_equation}) that
\be
\frac{\cot(a\sqrt{V_1})}{\sqrt{V_1}} \rightarrow 
\frac{\coth(b\sqrt{V_2})}{\sqrt{V_2}}\,.
\ee
Then, according to (\ref{INTEREST1_equation}), $T\rightarrow a$, 
i.e., there is no pulse separation. Conversely, when the separation
is increased, the difference between the areas of the positive and
negative pulses will also increase.

\item As $a\sqrt{V_1}$ approaches the maximum allowed value 
$\frac{\pi}{2}$ [c.f.~(\ref{AREAB_equation})], we note from 
(\ref{TANTANH_equation}) that $V_2 \rightarrow \infty$ and from
(\ref{INTEREST1_equation}) that $T\rightarrow a$. Thus, the 
quantum interest `rate' becomes infinite in this limit, suggesting
that it progressively becomes more and more difficult to reach 
this maximum amount of negative energy. The maximum value itself
is, of course, physically impossible to achieve.

\item If we take the limits $a,b\rightarrow0$ such that $a\rho_1$
and $b\rho_2$ remain finite, we would expect to recover the corresponding 
results for a pair of $\delta$-function pulses \cite{FT2}. If we set 
$B\equiv a\rho_1$ and $(1+\epsilon)B\equiv b\rho_2$, the constraint
(\ref{INTEREST1_equation}) implies that
\be
\epsilon \geq {6\pi BT\over1-6\pi BT}\,,
\ee
which was indeed the result obtained in \cite{FT2}.

\end{enumerate}

\newsection{Proof of quantum interest in two dimensions}

In this section, we shall present a general proof of the quantum interest
conjecture for minimally coupled, massless scalar fields in two dimensions. 
This will be done in two parts: firstly, to prove that the positive 
pulse must always over-compensate the negative pulse; and secondly, to 
show that the minimum amount of over-compensation increases with the 
pulse separation, and that there exists a (finite) maximum separation 
between the pulses.

The first part of the proof follows almost immediately from a theorem of 
Simon \cite{Simon},
concerning the existence of negative eigenvalues to the quantum-mechanical
Hamiltonian. In a form suitable to our purposes, it reads:

{\it Theorem\/}. Let $V(t)$ obey 
$\int_{-\infty}^\infty(1+t^{2})|V(t)|\,{\rm d}t
<\infty$, with $V(t)$ not a.e.\ zero\footnote{Almost everywhere (a.e.)
zero: zero everywhere except on a set of measure zero \cite{RS-I}.}. 
Then $H^{(1)}\equiv-{\rm d}^2/{\rm d}t^2+V(t)$ has a negative eigenvalue if 
\be
\int_{-\infty}^\infty V(t)\,{\rm d}t\leq0\,.
\ee

Since QI-compatibility is equivalent to the condition that there
are no negative eigenvalues to $H^{(1)}$, Simon's theorem implies 
that if $\int_{-\infty}^\infty\rho(t)\,{\rm d}t\leq0$, then $\rho(t)$ is not 
QI-compatible. In other words, QI-compatibility implies that the positive 
parts of an energy density must always over-compensate for any negative 
components. 

Simon's theorem is, in fact, far more general than what we need, since it is
valid for potentials (energy-density distributions) that have any number of 
positive and negative parts, and even for those that are not compactly
supported. In order to proceed, however, we shall begin assuming a more 
specific form of the energy distribution: namely, that it consists of a 
single continuous negative-energy part $V_{\rm gen}^-(t)$, followed 
after time $T$ by a single continuous positive-energy part 
$V_{\rm gen}^+(t)$. Furthermore, we demand that both $V_{\rm gen}^+(t)$ 
and $V_{\rm gen}^-(t)$ are compactly supported, i.e., 
they are each localised to a finite interval in time.

\begin{figure}[t]
\begin{center}
\includegraphics{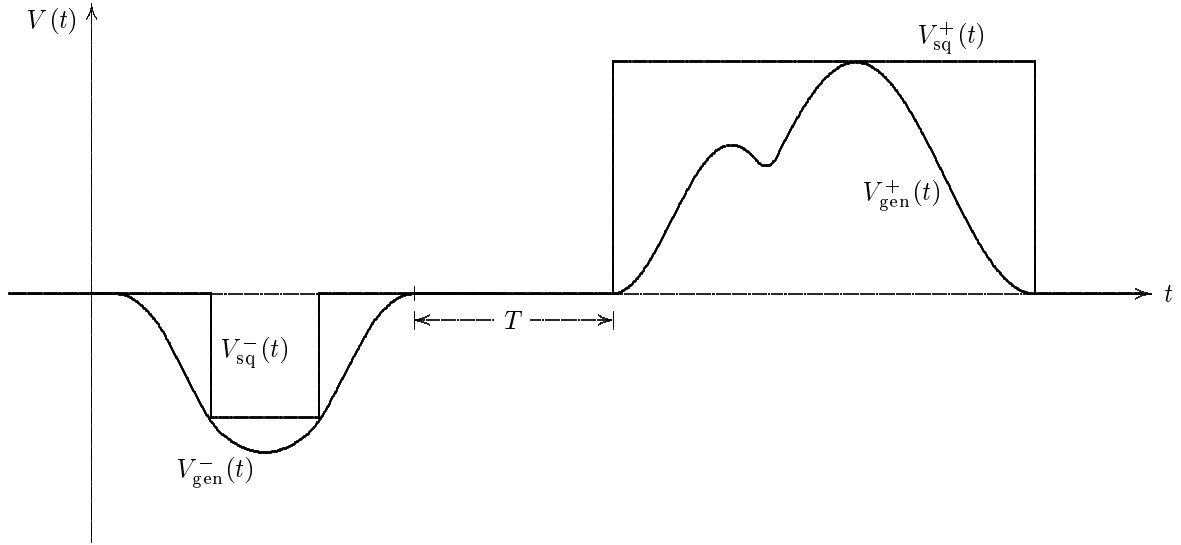}
\caption{Shape of a general potential $V_{\rm gen}(t)$, together with 
the corresponding square potential $V_{\rm sq}(t)$. The $\pm$ superscripts
denote their positive- and negative-energy parts, respectively. $T$ defines 
the separation between $V_{\rm gen}^-(t)$ and $V_{\rm gen}^+(t)$.}
\end{center}
\end{figure}

Given such an energy distribution 
$V_{\rm gen}(t)=V_{\rm gen}^+(t)+V_{\rm gen}^-(t)$, one can always 
construct a corresponding square-pulse distribution $V_{\rm sq}(t)$
of the type considered in the previous section, such that
\be
\label{min-max}
V_{\rm sq}(t)\geq V_{\rm gen}(t)
\ee
for all $t$. An example is illustrated in Fig.~4. The positive square pulse 
$V_{\rm sq}^+(t)$ is uniquely determined by the requirement that its 
height is equal to the maximum height of $V_{\rm gen}^+(t)$, and that 
its width coincides with the interval on which $V_{\rm gen}^+(t)$ is non-zero. 
The choice of the negative square pulse $V_{\rm sq}^-(t)$ is essentially 
free, but we shall choose it to be the one with the largest possible 
area that can `lie within' $V_{\rm gen}^-(t)$.

Because of the condition (\ref{min-max}), the min-max principle \cite{RS-IV}
can be used to show that the number of bound states of $V_{\rm sq}(t)$
is always less than or equal to the number of bound states of 
$V_{\rm gen}(t)$. This result is intuitively clear: the potential well 
of $V_{\rm sq}(t)$ is shallower, and the potential barrier higher, than 
that of $V_{\rm gen}(t)$, making the existence of bound states less 
likely for the former. QI-compatibility of $V_{\rm gen}(t)$ therefore 
implies the QI-compatibility of $V_{\rm sq}(t)$.

Now, suppose the height and width of $V_{\rm gen}^+(t)$ do not
increase as the separation $T$ between the pulses is increased. This implies
that the corresponding square-pulse potential $V_{\rm sq}(t)$ will also 
remain QI-compatible as $T$ is increased. However, we know from 
(\ref{INTEREST2_equation}) that this cannot be true; $V_{\rm sq}^+(t)$ 
has to increase in area as $T$ is increased. Thus we have a
contradiction.

A similar proof-by-contradiction can be used to show the existence of
a maximum separation $T$. Suppose $T$ can be made arbitrarily large, yet
a QI-compatible $V_{\rm gen}(t)$ is possible for some sufficiently large 
positive pulse. This implies that a QI-compatible $V_{\rm sq}(t)$ can
also be found. But we know from (\ref{PULSESEPB_equation}) that the 
latter is false; a maximum $T$ exists beyond which $V_{\rm sq}(t)$ is 
always QI-incompatible regardless of how large the positive pulse is. 

This completes the proof of the quantum interest conjecture.

We end with the remark that it is possible to obtain an upper bound on the 
maximal pulse separation for $V_{\rm gen}(t)$ using the results of Sec.~2, 
but, of course, it will not be optimal in general. This bound on $T$ is 
given by (\ref{INTEREST1_equation}), where $V_1$ and $a$ [$V_2$ and $b$] 
are the height and width of $V_{\rm sq}^-(t)$ [$V_{\rm sq}^+(t)$], 
respectively. In view of this, a tighter bound would be obtained if the
area of $V_{\rm sq}^-(t)$ was maximised (as was done above).

\newsection{Conclusion}

In this paper, the eigenvalue approach was used to study the problem
the quantum interest in two-dimensional minimally coupled, massless 
scalar field theory. The key to this approach lies in turning the 
problem into a more familiar and well-studied one, namely that of 
quantum mechanics on a line. We first explained in detail how the 
latter viewpoint can be used to investigate the case of double square 
energy pulses. We then generalised these results, using certain 
well-known properties of quantum-mechanical Hamiltonians, to obtain a 
proof of the quantum interest conjecture for general energy-density 
distributions.

The next obvious course of pursuit would be to extend the results of this
paper from two to four dimensions. The main difference is 
that $H^{(2)}$, given by (\ref{H}), is now a fourth-order differential 
operator, and so there is no longer a direct connection with the 
Hamiltonian in quantum mechanics. Still, it should be 
possible to consider specific energy-density distributions, such as 
that of double square pulses, by solving the appropriate fourth-order 
Schr\"odinger-type equation. The results should be interesting, 
especially in the light of the discovery in \cite{FT2} that the case 
of double $\delta$-function pulses in four dimensions can never be 
QI-compatible, no matter how close they are together. 

Ultimately, we would want to prove the quantum interest conjecture for
massless scalar fields in four dimensions. Since the existence of a 
maximum loan term has already been established in general \cite{FT2}, it 
remains to show that the interest `rate' is positive, perhaps using a 
similar strategy as in this paper. But to do so, a detailed 
study of the spectrum of fourth-order Schr\"odinger-type operators,
and in particular an appropriate generalisation of Simon's theorem
\cite{Simon}, is first needed.

\appendix

\newappendix{}

In this appendix, we show that $\frac{1}{k}\tanh^{-1} F(k)$ is a 
(branchwise) monotonically increasing function of $k$. It will be 
sufficient to consider only the case $F(k)\geq0$, although this 
result is generally valid.

Firstly, note that if we define $H(x)\equiv\tanh^{-1}x$ and $H'(x)\equiv
{\rm d}H(x)/{\rm d}x$, the inequality
\be 
x\geq\frac{H(x)}{H'(x)}
\ee
is satisfied for $0\leq x<1$. Setting $x=F(k)$, this can be used 
to show that
\ba
\frac{\rm d}{{\rm d}k}\left(\frac{1}{k}H(F(k))\right)
&\geq& H'(F(k)) \left(\frac{F'(k)}{k}-\frac{F(k)}{k^{2}}\right)\cr
&=& H'(F(k))\, \frac{\rm d}{{\rm d}k}\left(\frac{1}{k}F(k)\right).
\ea
Since $H'(F(k))>0$, the proof that  
$\frac{1}{k}\tanh^{-1}F(k)$ is monotonically increasing is reduced to 
showing that $\frac{1}{k}F(k)$ is monotonically increasing. 

Now, the derivative of $\frac{1}{k} F(k)$ is
\be
\label{deriv}
\frac{kV_1\omega a + k\big(V_1-2k^{2}\big) \sin \omega a \cos \omega a + \omega^{3} + 2\omega k^{2} \sin^{2}\omega a}{\omega^5\big(\sin \omega a - \frac{k}{\omega} \cos \omega a \big)^{2}}\,.
\ee
But the non-negativity of $F(k)$ implies that
\be
\big( V_1-2k^{2} \big) \sin \omega a \cos \omega a \geq \omega k \big( \cos^{2}\omega a - \sin^{2}\omega a \big)\,,
\ee
and this inequality can be used in (\ref{deriv}) to show that it is a 
positive function of $k$. We conclude that $\frac{1}{k} F(k)$, and therefore
$\frac{1}{k}\tanh^{-1} F(k)$, is a monotonically increasing function 
of $k$ when $0\leq F(k)<1$.

\newappendix{}

Given the function $G(k)$ as defined in (\ref{Gk}), we would like to show 
that $G(k)>G(0)$ for all $k\in(0,\sqrt{V_1})$. To this end, we shall 
instead prove that
\be
G(k) \geq \frac{\cot\omega a}{\omega} - \frac{\coth\lambda b}{\lambda} - (T-a)\,,
\ee
which is a slightly stronger result as the right-hand side is an increasing
function of $k\in(0,\sqrt{V_1})$, that is equal to $G(0)$ when $k=0$.

Since ${1\over k}F(k)>{1\over\omega}\cot\omega a$ for $0<\omega a<{\pi\over2}$, 
the problem is reduced to showing that
\be
\frac{g_{2}(k)}{g_{1}(k)}\leq k\left( T-a + \frac{\coth \lambda b }{ \lambda} \right).
\ee
This relation can be rewritten as
\ba
&&\lambda^{2} \left( k(T-a) \cosh k(T-a) - \sinh k(T-a) \right) \nonumber\\
&&{}+k^{2} \left( k(T-a) \sinh k(T-a) - \cosh k(T-a) \right) {} \nonumber\\
&&{}+\lambda k{\rm e}^{k(T-a)} \coth \lambda b \left( k(T-a) + \frac{k}{\lambda} \coth \lambda b\right) - \lambda k \coth \lambda b \sinh k(T-a) {}\nonumber\\
&&{}+\frac{k^{3}}{\lambda} \coth \lambda b \sinh k(T-a)\geq0\,.
\ea
Since the first and last terms on the left-hand side are non-negative, 
we only need to check that
\ba
&&k^{2} \left( k(T-a) \sinh k(T-a) - \cosh k(T-a) \right) {} \nonumber\\
&&{}+\lambda k{\rm e}^{k(T-a)} \coth \lambda b \left( k(T-a) + \frac{k}{\lambda} \coth \lambda b\right) - \lambda k \coth \lambda b \sinh k(T-a) \geq0\,.
\label{0001_equation}
\ea
Now, note that $x \sinh x - \cosh x \geq -1$, for any $x \geq 0$.
This implies that the left-hand side of (\ref{0001_equation}) is
\ba
&\geq&\lambda k{\rm e}^{k(T-a)} \coth \lambda b 
\left( k(T-a) + \frac{k}{\lambda} \coth \lambda b \right) 
- \lambda k \coth \lambda b \sinh k(T-a) - k^{2} \nonumber\\
&=&\lambda k \coth \lambda b 
\left( k(T-a){\rm e}^{k(T-a)} - \sinh k(T-a) \right) 
+ k^{2} \left({\rm e}^{k(T-a)} \coth^{2} \lambda b - 1 \right) \nonumber\\
&\geq&0\,,
\ea
where we have used the relations $\coth x \geq 1$,
$x{\rm e}^x - \sinh x \geq 0$, and ${\rm e}^x \geq 1$, for any 
$x \geq 0$, to obtain the final inequality.

\vspace{12mm}
\pagebreak[3]
\noindent{\bf Acknowledgements}
\nopagebreak
\medskip
\nopagebreak

We would like to thank C.~J.~Fewster for suggesting the proof in Appendix A, 
and for his comments on the manuscript.

\bigskip\bigskip

{\renewcommand{\Large}{\normalsize}
}

\begin{thebibliography}{99}

\bibitem{HE}
S.~W.~Hawking and G.~F.~R.~Ellis, 
{\it The large scale structure of space-time\/}
(Cambridge University Press, Cambridge, 1973), pp.~88--96.

\bibitem{Penrose}
R.~Penrose,
Phys.\ Rev.\ Lett.\  {\bf 14} (1965) 57.

\bibitem{MT}
M.~S.~Morris and K.~S.~Thorne,
Am.\ J.\ Phys.\  {\bf 56} (1988) 395.

\bibitem{MTY}
M.~S.~Morris, K.~S.~Thorne and U.~Yurtsever,
Phys.\ Rev.\ Lett.\  {\bf 61} (1988) 1446.

\bibitem{Ford:1990}
L.~H.~Ford and T.~A.~Roman,
Phys.\ Rev.\ D {\bf 41} (1990) 3662.

\bibitem{Ford:1992}
L.~H.~Ford and T.~A.~Roman,
Phys.\ Rev.\ D {\bf 46} (1992) 1328.

\bibitem{Alcubierre}
M.~Alcubierre,
Class.\ Quant.\ Grav.\  {\bf 11} (1994) L73
[arXiv:gr-qc/0009013].

\bibitem{Olum}
K.~D.~Olum,
Phys.\ Rev.\ Lett.\  {\bf 81} (1998) 3567
[arXiv:gr-qc/9805003].

\bibitem{Epstein} 
H.~Epstein, V.~Glaser and A.~Jaffe, 
Nuovo Cimento {\bf 36} (1965) 1016.

\bibitem{Tipler}
F.~J.~Tipler, 
Phys.\ Rev.\ D {\bf 17} (1978) 2521.

\bibitem{Klinkhammer}
G.~Klinkhammer,
Phys.\ Rev.\ D {\bf 43} (1991) 2542.

\bibitem{Ford}
L.~H.~Ford,
Phys.\ Rev.\ D {\bf 43} (1991) 3972.

\bibitem{Ford:1994}
L.~H.~Ford and T.~A.~Roman,
Phys.\ Rev.\ D {\bf 51} (1995) 4277
[arXiv:gr-qc/9410043].

\bibitem{Ford:1996}
L.~H.~Ford and T.~A.~Roman,
Phys.\ Rev.\ D {\bf 55} (1997) 2082
[arXiv:gr-qc/9607003].

\bibitem{Flanagan}
\'E.~\'E.~Flanagan,
Phys.\ Rev.\ D {\bf 56} (1997) 4922
[arXiv:gr-qc/9706006].

\bibitem{FE}
C.~J.~Fewster and S.~P.~Eveson,
Phys.\ Rev.\ D {\bf 58} (1998) 084010
[arXiv:gr-qc/9805024].

\bibitem{Pfenning:1996}
M.~J.~Pfenning and L.~H.~Ford,
Phys.\ Rev.\ D {\bf 55} (1997) 4813
[arXiv:gr-qc/9608005].

\bibitem{Pfenning:1997}
M.~J.~Pfenning and L.~H.~Ford,
Phys.\ Rev.\ D {\bf 57} (1998) 3489
[arXiv:gr-qc/9710055].

\bibitem{Pfenning}
M.~J.~Pfenning,
``Quantum inequality restrictions on negative energy densities in curved space-times,''
arXiv:gr-qc/9805037.

\bibitem{FT1}
C.~J.~Fewster and E.~Teo,
Phys.\ Rev.\ D {\bf 59} (1999) 104016
[arXiv:gr-qc/9812032].

\bibitem{Fewster:1999}
C.~J.~Fewster,
Class.\ Quant.\ Grav.\  {\bf 17} (2000) 1897
[arXiv:gr-qc/9910060].

\bibitem{Vollick:2000}
D.~N.~Vollick,
Phys.\ Rev.\ D {\bf 61} (2000) 084022
[arXiv:gr-qc/0001009].

\bibitem{Fewster:2001}
C.~J.~Fewster and R.~Verch,
Commun.\ Math.\ Phys.\  {\bf 225} (2002) 331
[arXiv:math-ph/0105027].

\bibitem{Pfenning:2001}
M.~J.~Pfenning,
Phys.\ Rev.\ D {\bf 65} (2002) 024009 [arXiv:gr-qc/0107075].

\bibitem{Ford:1999}
L.~H.~Ford and T.~A.~Roman,
Phys.\ Rev.\ D {\bf 60} (1999) 104018
[arXiv:gr-qc/9901074].

\bibitem{Pretorius}
F.~Pretorius,
Phys.\ Rev.\ D {\bf 61} (2000) 064005
[arXiv:gr-qc/9903055].

\bibitem{FT2}
C.~J.~Fewster and E.~Teo,
Phys.\ Rev.\ D {\bf 61} (2000) 084012
[arXiv:gr-qc/9908073].

\bibitem{Simon} 
B.~Simon, 
Annals Phys.\  {\bf 97} (1976) 279.

\bibitem{RS-IV}
M.~Reed and B.~Simon, {\it Methods of modern mathematical physics, IV. 
Analysis of operators\/} (Academic Press, New York, 1975), pp.~76--78.

\bibitem{GP}
A.~Galindo and P.~Pascual, {\it Quantum mechanics I\/} 
(Springer-Verlag, Berlin, 1990), pp.~132--133.

\bibitem{RS-I}
M.~Reed and B.~Simon, {\it Methods of modern mathematical physics, I. 
Functional analysis\/} (Academic Press, New York, 1980), p.~17.


\end{thebibliography}
\end{document}